# Quantifying the effect of electrical conductivity on the rate-performance of nanocomposite battery electrodes

Ruiyuan Tian,[1,2] Nolito Alcala,[3] Stephen JK O'Neill,[1,2] Dominik Horvath,[1,2] João Coelho,[2,4] Aideen Griffin,[1,2] Yan Zhang,[5] Valeria Nicolosi,[2,4] Colm O'Dwyer,[2,5] Jonathan N Coleman[1,2]*

[1]School of Physics, Trinity College Dublin, Dublin 2, Ireland

[2]AMBER Research Center, Trinity College Dublin, Dublin 2, Ireland

[3]School of Physics, Technological University Dublin, City Campus, Kevin Street, D08 NF82, Ireland

[4]School of Chemistry, Trinity College Dublin, Dublin 2, Ireland

[5]School of Chemistry, University College Cork, Tyndall National Institute, and Environmental Research Institute, Cork T12 YN60, Ireland

*colemaj@tcd.ie (Jonathan N. Coleman); Tel: +353 (0) 1 8963859.

ABSTRACT: While it is well-known that electrode conductivity has a critical impact on rate-performance in battery electrodes, this relationship has been quantified only by computer simulations. Here we investigate the relationship between electrode conductivity and rate-performance in Lithium-Nickel-Manganese-Cobalt-Oxide (NMC) cathodes filled with various quantities of carbon black, single-walled carbon nanotubes and graphene. The electrode conductivity is always extremely anisotropic with the out-of-plane conductivity, which is most relevant to rate-performance, roughly ×1000 smaller than the in-plane conductivity. For all fillers the conductivity increases with filler loading although the nanotube-filled electrodes show by far the most rapid increase. Fitting capacity versus rate curves yielded the characteristic time associated with charge/discharge. This parameter increased linearly with the inverse of the out-of-plane conductivity, with all data points falling on the same master curve. Using a simple mechanistic model for the characteristic time, we develop an equation which matches the experimental data almost perfectly with no adjustable parameters. This implies that increasing the electrode conductivity improves the rate-performance by decreasing the RC charging time of the electrode. This model shows the effect of electrode resistance on the rate-performance to become negligible in almost all cases once the out-of-plane conductivity of the electrode exceeds 1 S/m. Our results show that this can be achieved by including <1wt% single-walled carbon nanotubes in the electrode.

INTRODUCTION

Rechargeable batteries based on the storage of Lithium ions are becoming more and more important for many applications including electric vehicles, mobile electronics and even large-scale energy storage.[1,2] While much of the focus within the research community has been on maximising capacity and energy density, somewhat less attention has been given to optimising rate-performance.[3] Nevertheless, achieving high rate-performance is critical to achieving rapid charging or high power delivery in a range of applications.[4]

It is well known that many factors affect the rate-performance of an electrode/electrolyte system, including the solid-state diffusion time, the time taken for ions to diffuse within the electrolyte and the ability of the electrode material to rapidly distribute charge.[5-9] This latter factor generally requires intervention as many battery materials have relatively low electrical conductivity. To address this, conductive additives are almost always incorporated into the electrode to reduce electrode resistance.[10] In most cases, tried and tested formulations are used, with the addition of ~10wt% carbon black being particularly common.

However, there does not seem to be a clear rule defining the aims associated with incorporating the conductive additives. For instance, it would be useful to know exactly what minimum conductivity is being targeted. This would allow one to minimise the conductive additive content, thus maximising the active material content, while still reaching the target conductivity. In addition, the literature does not generally contain much discussion as to what aspect of conductivity is important. For example, films containing networks of conducting nano-carbons, especially those cast from liquids, can be very anisotropic, leading to significant differences between in-plane and out-of-plane conductivity.[11,12] Although the in-plane conductivity is easy to measure and is often reported,[13,14] the out-of-plane conductivity is probably more relevant in battery electrodes as it governs transport of charge from current collector to active sites.[9] Indeed, it has been shown that the out-of-plane conductivity corresponds very well to the electrode resistance measured by impedance spectroscopy.[15] However, we are aware of no quantitative examination of the relationship between either in-plane or out-of-plane conductivity and rate-performance. Such a relationship would be extremely useful as it would allow the identification of the minimum conductivity required to optimise rate-performance.

In this work we study both the in-plane and out-of-plane conductivities of battery electrodes based on NMC filled with three different fillers, carbon black (CB), single-walled carbon nanotubes (CNT) and graphene (Gra), at various loadings. While we find both in-plane and out-of-plane conductivities to scale with filler volume fraction as per percolation theory, the out-of-plane conductivity was roughly three orders of magnitude lower that that measured in-plane. Rate measurements showed the characteristic time associated with charge/discharge to scale inversely with out of plane conductivity. Using a simple mechanistic model, we can match the data almost perfectly with no adjustable constants. Then, using the model, we show rate-performance optimisation to occur in almost all circumstances once the out-of-plane conductivity exceeds 1 S/m.

RESULTS AND DISCUSSION

*Electrode conductivity*

We produced a range of electrodes of NMC loaded with varying mass fractions ($M_f$) of CNTs (0.01%<$M_f$<4%), graphene (0.1%<$M_f$<30%) and carbon black (0.1%<$M_f$<20%). In each case, we were careful to measure areal mass (M/A) and thickness ($L_E$) of the electrodes, allowing us to calculate their density ($\rho_E$) and porosity ($P_E$) as well as filler volume fraction ($\phi$). In general, the electrodes were ~100 μm thick. SEM images (figure 1) show the electrodes to consist of loosely packed disordered arrays of NMC particles (diameter ~5-10 μm) surrounded by a lose network of filler particles. Such a system where the matrix (i.e. active) particles are larger than the filler (i.e. CB, CNT or graphene) particles is called a segregated network and has been shown to result in high conductivities at relative low filler mass fractions.[15,16]

For each electrode, we measured both the in-plane (IP) and out-of-plane (OOP) apparent conductivity using the two-probe technique as described in methods. We use the term apparent conductivity as two probe techniques include the effects of contact resistance which can have a significant impact when the material resistance is small. While contact resistance effects can be removed by using 4-probe measurements, this is not straightforward for OOP measurements. In composites, conductivities are usually analysed in terms of filler volume fraction, $\phi$, rather than $M_f$. The volume fraction can be calculated from $\phi = M_f \rho_E / \rho_{filler}$, which is found by defining $\phi = V_{filler} / V_{electrode}$.[12]

Shown in figure 2A-C are both in-plane ($\sigma_{IP}$) and out-of-plane ($\sigma_{OOP}$) conductivities for composites with CB (A), graphene (B) and CNT fillers (C), all as a function of $\phi$. In each case, the conductivity increases rapidly with $\phi$, once a minimum filler volume fraction had been surpassed. For all materials, the maximum IP conductivity was ~1000 S/m while the highest OOP conductivity observed was ~0.1 S/m. Over all filler loadings, the in-plane conductivity was between ×4 and ×3000 larger than $\sigma_{OOP}$. Such large conductivity anisotropies have been observed before for nanostructured networks[11,17] and occur when the networks are partially aligned in the plane of the film. Such conductivity anisotropy will have significant implications for performance in battery electrodes because measurement of in-plane conductivity will significantly over-estimate the effect of the conductivity on rate-performance. Unusually for such composites, both in-plane and out-of-plane conductivities saturated for the nanotube filled composites as the volume fraction surpassed ~0.2%.

For composites filled with conductive additives, the conductivity is described by percolation theory. Within this model, the conductivity increases only above a critical volume fraction where the first complete conductive path is formed, a value known as the percolation threshold, $\phi_c$. Above this threshold, the composite conductivity, $\sigma$, scales as:

$$\sigma = \sigma_0 (\phi - \phi_c)^t \qquad (1)$$

where $\sigma_0$ is a constant related to the conductivity of the filler network and $t$ is the percolation exponent.[12,18]

With this in mind, it is clear that the percolation threshold for CB and graphene composites is ~1-2vol% but much lower for the nanotube-filled composites. Equation (1) fits the data extremely well for the CB and graphene composites in both IP and OOP directions with all fit parameters given in table 1. The percolation thresholds are very similar between IP and OOP directions indicating that network connectivity is similar in the in-plane and out-of-plane directions. The in-plane percolation exponents were close to the universal, 3-dimensional value of 2.0 which indicates that the distribution of inter-particle junction resistances is fairly narrow.[12] However, OOP exponents were slightly lower, perhaps due to network alignment effects.[19] However, the major difference between IP and OOP parameters for the CB and graphene samples were the $\sigma_0$ values which were approximately ×1000 higher in the IP direction. Such a large anisotropy confirms that the conducting networks are significantly aligned in the plane of the electrode.

However, the σ-ϕ curves for the nanotube-filled samples behaved differently, saturating at higher loadings. We can explain this by noting that, for two-probe measurements, the measured resistance is the sum of composite and contact resistance ($R_C$). Converting these resistances to conductivity via the electrode area, A, and separation, L, yields the effective (ie measured) conductivity

$$\sigma_E = \frac{1}{R_C A / L + 1/\sigma} \tag{2a}$$

where σ is the composite resistance. We can use equation (1) to replace σ, yielding:

$$\sigma_E = \frac{1}{R_C A / L + \left[\sigma_0 (\phi - \phi_c)^t\right]^{-1}} \tag{2b}$$

We note that when the contact resistance is very small, equation (2b) reverts to equation (1). We find equation (2b) fits the data extremely well. As shown in table 1, the exponents are very similar to the other materials. However, the percolation thresholds are considerably lower while the $\sigma_0$ values are much higher than the CB and graphene-based composites. Taken together, this means carbon nanotubes yield much higher conductivities at lower loading levels compared to other fillers. In addition, we use the fits to estimate the contact resistances, $R_C A$, for the CNT-filled composites. These work out to be $9 \times 10^{-6}$ $\Omega m^2$ and $3.3 \times 10^{-4}$ $\Omega m^2$ for the IP and OOP directions. This difference is to be expected based on the nature of the contacts (see methods), with the top contact in the OOP measurement being relatively poorly connected to the electrode.

Once we know $\sigma_0$, $\phi_c$ and n for the CNT-filled electrodes, it is possible to estimate the composite conductivity (i.e. neglecting contact effects) as a function of volume fraction, ϕ, using equation (1). We have plotted equation (1) on figure 2C (solid lines) using the fit parameters given in table 1. These curves confirm that the true composite conductivities for nanotube-filled electrodes can be significantly higher than for the other systems.

*Measuring rate performance*

This work shows clearly that the out-of-plane conductivity of these electrodes is significantly lower than the in-plane conductivity and varies greatly depending on filler. Because it controls transport of charge from current collector to Li storing sites, we would expect the OOP conductivity to directly impact the electrodes rate-performance. As a result, it is worth measuring capacity-rate data for each composite type at a number of filler loadings with the aim of correlating rate-performance with $\sigma_{OOP}$. To do this we fabricated electrodes based on NMC with various mass fractions of CB, graphene and CNT in the same way as described above. First, we performed galvanostatic charge discharge (GCD) measurements to check the electrodes were performing correctly. Shown in figure 3 are selected second cycle charge/discharge curves for three mass fractions for each composite type. In all cases, the GCD curves are consistent with previous reports[20] while the capacity increasing with filler loading as expected.[13] Normally, rate-performance measurements are carried out by performing GCD measurements at a range of currents. However, such measurements are prohibitively slow, especially when many samples are being measured and enough different rates to perform quantitative analysis are required. To get around this problem, we used a recently reported, relatively rapid method of making rate measurements: chronoamperometry (CA).

Heubner et al.[21] have shown that CA is a very effective technique for performing rate-performance measurements. This method has the advantages that it is quicker than GCD and yields many more data points down to lower rates. In practice, a potential step is applied to the electrode and the current transient measured. Heubner et al. published equations to transform the I(t) data into capacity as a function of C-rate.

However, we have previously argued that quantitative analysis of rate-performance measurements is better performed on plots of capacity versus charge/discharge rate, R where R is defined as[9]

$$R = \frac{I/M}{(Q/M)_E} \qquad (3)$$

where $(Q/M)_E$ is the measured experimental capacity, rather than the theoretical value. In this way, R is related to the actual charge/discharge time. We have shown that the CA current transient can be converted to specific capacity, Q/M, and R using:[22]

$$R = \frac{I(t)/M}{\int_0^t (I(t)/M)dt} \qquad (4a)$$

and

$$Q/M = \int_0^t (I(t)/M)dt \qquad (4b)$$

We have shown that these equations give capacity-rate curves which match extremely well to those obtained by GCD.[22] However, they can be measured in approximately one third of the time.

Presented in figure 4 are Q/M vs. R curves for each of the three composite types for three different mass fractions. The first thing to note is that the CA derived curves have the same shape as standard GCD-derived rate curves. The main difference is the much higher data density. Secondly, these curves clearly show both the rate-performance and the low-rate capacity to increase with filler loading as expected.

*Fitting rate data*

In order to quantitatively analyse the relationship between rate-performance and electrode conductivity, it is necessary to extract a number from each capacity-rate graph which quantifies the rate-performance. Recently,[9] we proposed a semi-empirical equation for fitting capacity-rate data which outputs three fit parameters to assess rate-performance:

$$\frac{Q}{M} = Q_M \left[ 1 - (R\tau)^n \left( 1 - e^{-(R\tau)^{-n}} \right) \right] \quad (5)$$

Here $Q_M$ is the specific capacity at very low rate, $\tau$ is a time constant associated with charge/discharge and is a measure of the rate at which $Q/M$ starts to fall off.[9,23] This parameter is particularly important as low time constants mean fast charge/discharge and indicate good rate-performance. Finally, $n$ is an exponent describing how rapidly $Q/M$ decays at high rate with diffusion-limited electrodes showing $n\sim0.5$ while capacitive-limited (i.e electrically limited) electrodes yield $n\sim1$.[9] Knowledge of $n$ and especially $\tau$ allows a proper, quantitative assessment of the rate-performance of a given electrode.

We have used equation (5) to fit all of our Q/M vs. R curves with examples of fits shown in figure 4. In all cases the fits were very good giving us confidence in the accuracy of the fit values. These fit values are plotted versus filler mass fraction in equation (5) for each composite type. While $Q_M$ is not an indicator of rate-performance, we plot it in figure 5A to confirm the results to be as expected. In line with previous results, we find the low-rate capacity to increase with mass fraction of conductive filler.[13] Interestingly, the capacity increases occur at much lower mass fractions for the nanotube-filled samples compared to the CB- and graphene-filled electrodes. The exponent, n, is plotted versus $M_f$ in figure 5B. For low mass fractions, n is closer to 1 than 0.5 in all cases, consistent with these electrodes being predominately electrically limited (i.e. limited by the RC charging time of the electrode).[9,22] However, in each case, n appears to fall slightly with filler loading. This is consistent with increasing conductivity reducing the resistance of the system, thus slightly shifting the rate limiting effect from electrically- to diffusion-limited.[9]

However, most important for rate-performance is the characteristic time, τ. This parameter is a measure of the rate, above which capacity begins to fall off. As such it can be thought of as approximately the minimum charge/discharge time where the full low-rate capacity can be achieved. As such, this parameter nicely quantified rate-performance with better performance associated with low τ. As shown in figure 5C, in all cases τ falls with filler $M_f$, behaviour which has been observed previously.[9] Interestingly, the nanotube-filled composites reach lower values of τ at much lower loadings compared to the other two materials. This would suggest carbon nanotubes to be the best fillers when rate-performance is concerned.

*Mechanistic analysis*

We can understand these results by considering a model which we recently reported that describes τ in terms of the various timescales associated with ion motion in the system.[9] There are three main contributions to τ: the RC time constant of the system, the timescale associated with diffusion and the time associated with the electrochemical reaction. Each of these contributions can be broken into one or more terms within the equation which we number below. The RC terms include contributions from the electrical resistance of the electrode (1) as well as the ionic resistances of the electrolyte within the pores of the electrolyte (2) and within the separator (4). The diffusive terms include contributions from the times required for ions to diffuse through the electrolyte-filled porous interior of the electrode (3), the time required to diffuse through the separator (5) and the solid-state diffusion time (6). The final term (7) described the timescale associated with the electrochemical reaction, $t_c$.

This yields the following equation[9]

$$\tau = L_E^2 \left[ \frac{C_{V,eff}}{2\sigma_{OOP}} + \frac{C_{V,eff}}{2\sigma_{BL}P_E^{3/2}} + \frac{1}{D_{BL}P_E^{3/2}} \right] + L_E \left[ \frac{L_S C_{V,eff}}{\sigma_{BL}P_S^{3/2}} \right] + \left[ \frac{L_S^2}{D_{BL}P_S^{3/2}} + \frac{L_{AM}^2}{D_{AM}} + t_c \right] \quad (6a)$$

Term      1      2      3      4      5      6     7

Here $C_{V,eff}$ is the effective volumetric *capacitance* of the electrode, $\sigma_{OOP}$ is the out-of-plane electrical conductivity of the electrode material, $P_E$ and $P_S$ are the porosities of the electrode and separator respectively while $L_S$ is the separator thickness. Here $\sigma_{BL}$ is the overall (anion and cation) conductivity of the bulk electrolyte (S/m) while $D_{BL}$ is the ion diffusion coefficient in the bulk electrolyte. In addition, $L_{AM}$ is the solid-state diffusion length associated with the active particles (related to particle size); $D_{AM}$ is the solid-state Li ion diffusion coefficient within the particle. We note that the volumetric capacitance of a battery electrode may not be known. However, we have shown empirically that $C_{V,eff}$ is directly proportional to the low-rate volumetric capacitance of the electrode, $Q_V$, such that: $C_{V,eff}/Q_V = 28$ F/mAh.[9]

Here, we are interested in the dependence of $\tau$ on $\sigma_{OOP}$. In figure 6A we plot $\tau$ versus $\sigma_{OOP}$ for all three materials. We note that for the high CNT loading levels, we used figure 1, combined with the percolation fit parameters, to estimate the composite conductivity, removing the contribution of contact resistance. We find $\tau$ to fall significantly with increasing $\sigma_{OOP}$.

We can understand this behaviour by combining equation (6a) with the empirical relationship between $C_{V,eff}$ and $Q_V$ yielding:

$$\tau = \frac{14 Q_V L_E^2}{\sigma_{OOP}} + \beta \tag{6b}$$

where $\beta$ is just shorthand for terms 2-7 and $Q_V$ should be expressed in mAh/m$^3$. This equation implies that $\tau$ should scale linearly with $1/\sigma_{OOP}$. As shown in figure 6A, we find this relationship to describe the data reasonably well, albeit with some scatter.

We should not be surprised that the data in figure 6A is slightly scattered because, as shown in figure 5A, $Q_M$ shows a non-trivial variation over the samples. This means $Q_V$, which appears in equation (6b) will also vary (because $Q_V = \rho_E Q_M$). In addition, there are small unavoidable variations in the electrode thickness, $L_E$, over the samples. To combat these problems, we rearrange equation (6b) slightly to read.

$$\frac{\tau}{L_E^2} = 14 \frac{Q_V}{\sigma_{OOP}} + \frac{\beta}{L_E^2} \tag{7}$$

This implies that a graph of $\tau / L_E^2$ vs. $Q_V / \sigma_{OOP}$ should be a straight line with a slope which is material independent at 14 F/mAh. To plot this graph, we use our electrode density measurements to calculate $Q_V$ for each electrode. This graph is presented in figure 6B and shows a very well-defined straight line with reduced scatter compared to the data in figure 6A.

To test for quantitative agreement, we do not fit the data using equation (7). Instead, we directly plot equation (7) on figure 6B. The model predicts that the slope of this plot be 14 F/mAh while the intercept, $\beta / L_E^2$, can be found by using reasonable values of the electrode parameters in terms 2-7 of equation (6a). The parameters used are given in the caption of figure 6B (and are justified in the SI) and yield a value of the intercept to be $\beta/L_E^2 = 3.5\times 10^{10}$ s/m$^2$. Plotting equation (7) using this slope and intercept gives the solid line in figure 6B.

We find the agreement between the plot of equation (7) and the data in figure 6B remarkable. Such agreement between experiment and theory has a number of implications. First it strongly supports the validity of the model represented by equation (6a). This is important as it gives us confidence that the model can be used to analyse data or to predict behaviour. Secondly, the slope of equation (7) is determined by the empirical relationship between electrode volumetric capacity and volumetric capacitance reported in ref[9]. The almost perfect match between the slopes of model and data in figure 6B strongly supports this empirical relationship. Finally, this data confirms that it is the out-of-plane electrode conductivity that determines rate behaviour (rather than $\sigma_{IP}$).

*Predicting minimum required electrode conductivities.*

The data in figure 6A suggests that, at least for the electrodes under study here, the time constant is minimised once $\sigma_{OOP}$ exceeds about 1 S/m. This occurs because once $\sigma_{OOP}$ gets large enough, term 1 in equation (6a) becomes negligible compared to the rest of the terms. We can use this idea to identify the minimum electrode conductivity required to render term 1 negligible for any electrode. We can do this by imposing the (somewhat arbitrary) condition that term 1 becomes unimportant when it falls below 10% of the sum of the other 6 terms. Expressing this condition and then rearranging gives an expression for the minimum out-of-plane conductivity required to optimise rate-performance (with respect to filler content) by eliminating term 1:

$$\sigma_{OOP,Min} = \frac{14Q_V}{0.1\left[\dfrac{14Q_V}{\sigma_{BL}P_E^{3/2}} + \dfrac{1}{D_{BL}P_E^{3/2}} + \dfrac{28Q_V L_S/L_E}{\sigma_{BL}P_S^{3/2}} + \dfrac{L_S^2/L_E^2}{D_{BL}P_S^{3/2}} + \dfrac{\tau_{SSD}+t_c}{L_E^2}\right]} \qquad (8)$$

We note that, for reasons which will become clear, in equation (8) we have combined $L_{AM}$ and $D_{AM}$ in terms of the solid-state diffusion time, $\tau_{SSD} = L_{AM}^2/D_{AM}$. Of the parameters within equation (8) the only ones that can vary significantly (i.e. by orders of magnitude) between electrodes are $Q_V$, $L_E$ and $\tau_{SSD}$ and $t_c$. Typical values of $Q_V$ vary between tens and thousands of mAh/cm$^3$ depending on the material, while the majority of electrodes would have thicknesses between 1 µm and a maximum of ~1 mm.[15] By analysing a large number of published papers, we recently showed that $\tau_{SSD}$ tends to fall in the range 1-10$^4$ s.[24] Finally, while $t_c$ can be hard to pin down, Jiang et al.[6] have discussed values from 0.1-200 s. Of the other parameters, in real electrodes, the porosity tends to occupy a relative narrow range between ~0.4 and 0.6,[25] while the other parameters have reasonably standard values: $\sigma_{BL}$~0.5 S/m, $D_{BL}$~3×10$^{-10}$ m$^2$/s, $P_S$~0.4, $L_S$~25 μm (although here, $L_S$=16 μm).[9]

To estimate the minimum conductivity required to optimise rate-performance, we use equation (8) to plot $\sigma_{OOP,Min}$ in figure 7 as a contour plot versus $Q_V$ and $L_E$ using $Q_V$- and $L_E$-ranges as described above. We use the values of $\sigma_{BL}$, $D_{BL}$, $P_S$ and $L_S$ given above and take $P_E$=0.5. We plot two separate graphs, each for different values of the combination of $\tau_{SSD} + t_c$. Considering the numbers above, we take maximum and minimal values of $\tau_{SSD} + t_c$ of $10^4$ and 1 s respectively. These graphs clearly show that under almost any circumstances, an out-of-plane conductivity of 1 S/m will be enough to render term 1 in equation (6a) negligible, and thus optimise rate-performance from a filler perspective. With reference to figure 2, attaining $\sigma_{OOP}$=1 S/m would require >10vol% (i.e. >12wt%) CB or graphene but <1vol% (<1.3wt%) carbon nanotubes. This shows that carbon nanotubes have significant advantages as conductive additives in battery electrodes.

CONCLUSIONS

In this work we have shown that composite battery electrodes of NMC filled with three different conductive additives, carbon black, graphene or carbon nanotubes, show significant conductivity anisotropy, with out-of-plane conductivities ($\sigma_{OOP}$) roughly ×1000 lower than those measured in-plane. While carbon black or graphene loadings of >10wt% are required to reach OOP conductivities of 1 S/m, this level can be achieved with ~1wt% of carbon nanotubes. We found the rate-performance of such composite electrodes to depend strongly on filler loading. By fitting capacity-rate curves to an empirical equation, we extracted the characteristic charge/discharge time, τ, for each electrode. Informed by a simple mechanistic model, we found τ to scale approximately linearly with 1/$\sigma_{OOP}$ for all materials. By plotting $\tau / L_E^2$, where $L_E$ is the electrode thickness, versus $Q_V / \sigma_{OOP}$, where $Q_V$ is the electrode volumetric capacity, we found all data to collapse onto a linear master curve. This curve agreed almost perfectly with the predictions of the model with no adjustable fitting parameters. This allows us to use this model to estimate a minimum out-of-plane conductivity of 1 S/m required to optimise rate-performance.

This work highlights the importance of the out-of-plane conductivity to rate-performance in batteries and shows that conductivity measured in-plane is not a good metric for battery performance. It also shows that the loading level required to achieved sufficient conductivity varies very strongly with filler content, with carbon nanotubes showing the greatest efficiency in this regard. Finally, we emphasise that simple mechanistic models can accurately predict experimental data without the need to perform complex simulations.

METHODS

Samples for in-plane and out-of-plane conductivity measurements were prepared via the conventional slurry-casting method. $LiNi_{0.8}Co_{0.1}Mn_{0.1}O_2$ (NMC811) powder (MTI Corporation) was mixed with the respective conductive additive: CNTs (0.2 wt% CNT in NMP, 2wt% PVDF as a surfactant stabilizer, Tuball, OCSiAl), CB (Timical Super C65, MTI Corp.), Graphene (Graphene Powder, Tianyuan Empire), Polyvinylidene Fluoride (PVDF, EQ-Lib-PVDF, MTI Corp) and with sufficient amounts of N-Methyl-2-pyrrolidone (NMP) to form the slurry. There was 10wt% PVDF in most of the samples. However, the PDVF loading was increased to improve the Critic Crack Thickness (CCT)[15] of samples with extremely high loading of CB (18wt% and 22wt% PVDF in 15wt%CB and 20wt%CB samples). Samples for in-plane measurements were adhered to glass slides, whereas out-of-plane samples were cast onto an Al current collector using a doctor blade. All samples were dried overnight at 40 °C while the mass loading of active material (NMC811) was kept roughly constant at ~15 mg cm$^{-2}$.

Each in-plane sample was cut into a rectangular shape and silver wires were attached to the ends of the samples by painting them on with silver paint. This configuration allowed for intimate contact between sample and probe and in-plane conductivities were measured using the 2-point probe method. As for out-of-plane conductivity, circular disc electrodes with diameter = 12 mm were prepared by using a coin-cell disc puncher. Each electrode was then assembled into 2032-type coin cells in an Ar-filled glovebox (UNIlab Pro, Mbraun) in the following geometry: top, spring, two spacers, electrode, current collector, bottom. Out-of-plane conductivities were then measured using the two-point probe method. We expect the contact resistance between the top conductive spacer and the electrode to be non-trivial.

The electrochemical properties of the electrodes were measured in half cell (PAT-cell, EC Lab, BioLogic). All coin cells were assembled in an Ar-filled glovebox (UNIlab Pro, Mbraun). The dried electrodes were cut into 12 mm diameter discs and paired with Li metal discs (diameter= 16 mm). Celgard 2032 (thickness = 16 µm) was used as a separator. The electrolyte was 1.2 M $LiPF_6$ dissolved in EC/EMC (1:1 in v/v, BASF) with 10wt% Fluoroethylene carbonate (FEC).

The tests were performed at a potentiostat (VMP3, Biologic). The GCD measurements (at I/A=17 mA/g) were performed for 2-3 cycles to form stable solid-electrolyte- interface (SEI) film in the half cells, and the voltage range was 3−4.3 V. After the capacities were stable, the cells were charged at I/A=17 mA/g to 4.3 V, and CA measurements were performed for discharge at 3 V.[22]

**Acknowledgments:** The authors acknowledge the SFI-funded AMBER research centre (SFI/12/RC/2278) and Nokia for support. JNC thanks Science Foundation Ireland (SFI, 11/PI/1087) and the Graphene Flagship (grant agreement n°785219) for funding.

FIGURES and TABLES

|  | In plane | Out of plane |
|---|---|---|
| Carbon black | | |
| $\sigma_0$ (S/m) | $1.35 \times 10^4$ | 4.50 |
| $\phi_c$ (vol%) | 0.9 | 0.7 |
| t | 1.99 | 1.5 |
| Graphene | | |
| $\sigma_0$ (S/m) | $3.66 \times 10^4$ | 7.79 |
| $\phi_c$ (vol%) | 2.3 | 2.1 |
| t | 2.11 | 1.71 |
| CNTS | | |
| $\sigma_0$ (S/m) | $1.35 \times 10^8$ | $6.2 \times 10^4$ |
| $\phi_c$ (vol%) | 0.01 | 0.01 |
| t | 2.0 | 1.87 |

Table 1: Percolation fit parameters found by fitting the data in figure 2 using equation (1).

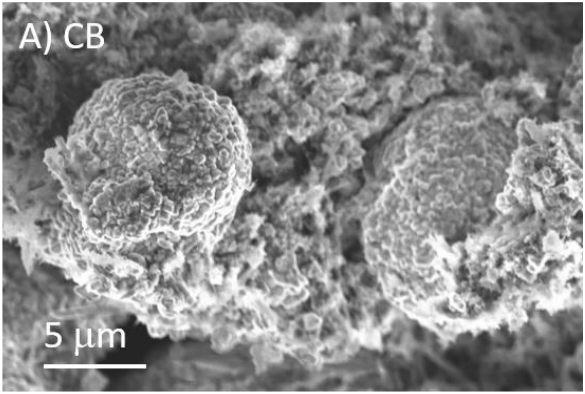

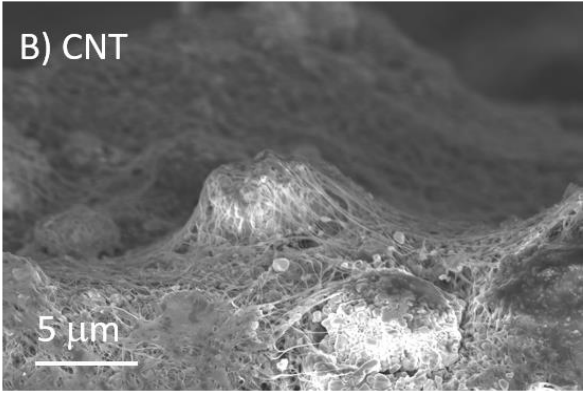

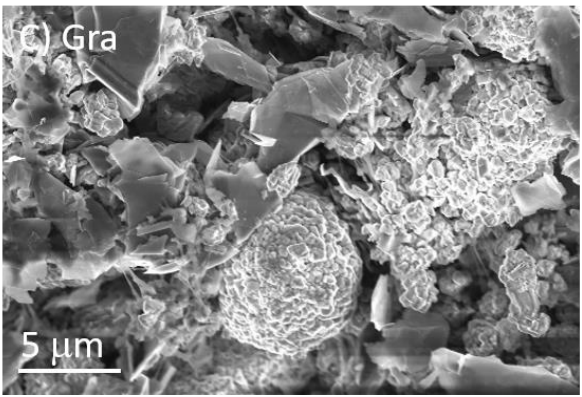

Figure 1: SEM images of fracture surfaces for 6wt% CB, 1wt% CNT, and 10wt% Graphene of each composite type.

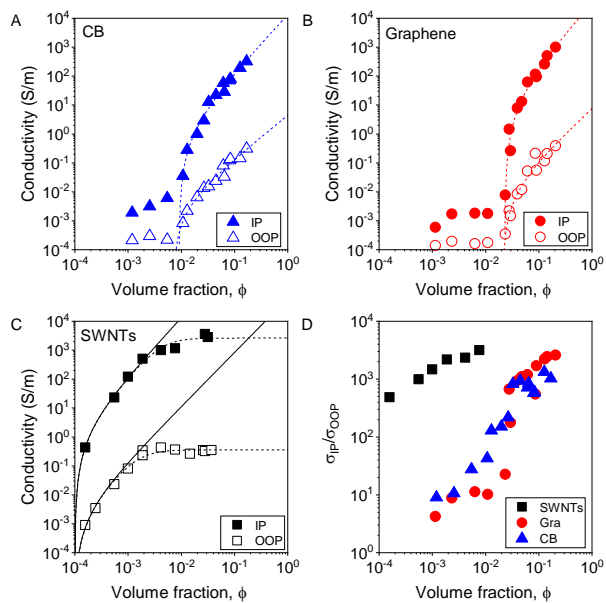

Figure 2: Conductivity of composite electrodes based on NMC811 filled with various conductive fillers. A-C) Measured conductivity as a function of volume fraction of conductive additives for composites filed with carbon black (A), graphene (B) and carbon nanotubes (C). The open symbols represent out-of-plane conductivity while the solid symbols represent in-plane conductivity. In A-C the dashed lines represent percolation fits (equation (1)). In C, the fits include the effect of contact resistance (equation (2b)). The solid lines represent the conductivity, estimated from the fits with the effect of contact resistance removed (i.e. using equation (1)). D) Ratio of in-plane to out of plane conductivity plotted versus volume fraction.

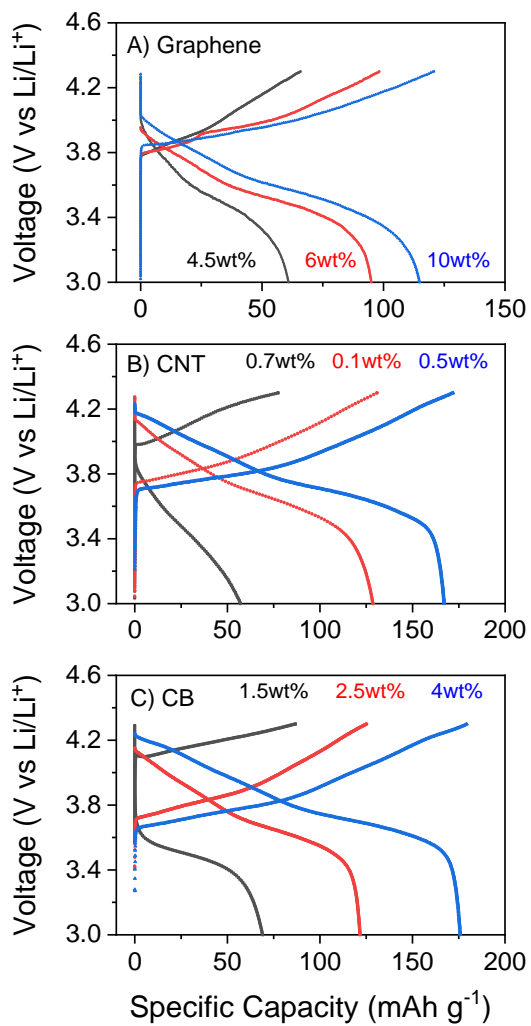

Figure 3: Second cycle charge-discharge curves for electrodes of NMC811 filled with different loadings of A) graphene, B) carbon nanotubes and C) carbon black. All measurements were performed at I/A=17 mA/g.

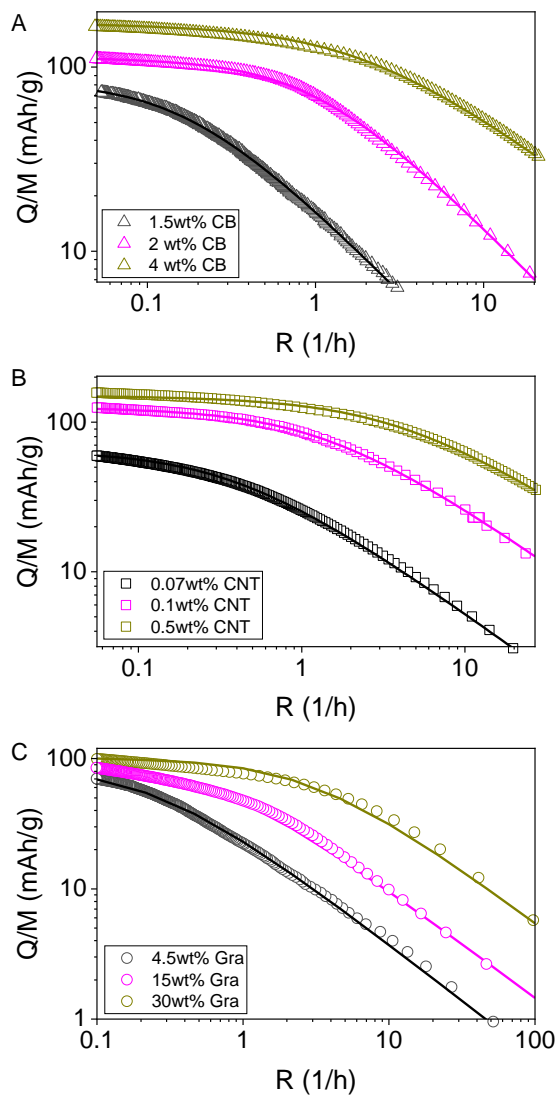

Figure 4: Rate-performance data for NMC811 electrodes incorporating carbon based conductive additives. A-C) Specific capacity (normalised to active mass) plotted versus rate NMC811-based electrodes filled with various loadings of carbon black (A), single-walled carbon nanotubes (B) and graphene (C).

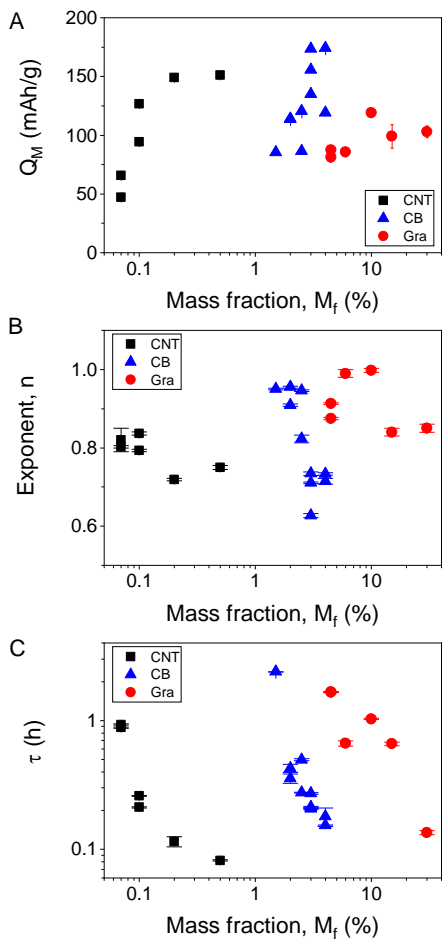

Figure 5: Rate-data fit parameters as a function of mass fraction of conductive additive. A) Low-rate specific capacity, $Q_M$, (B) characteristic time, $\tau$, (C) and rate exponent, $n$, each plotted against mass fraction for all three types of conductive additive (single-walled carbon nanotubes, carbon black and graphene).

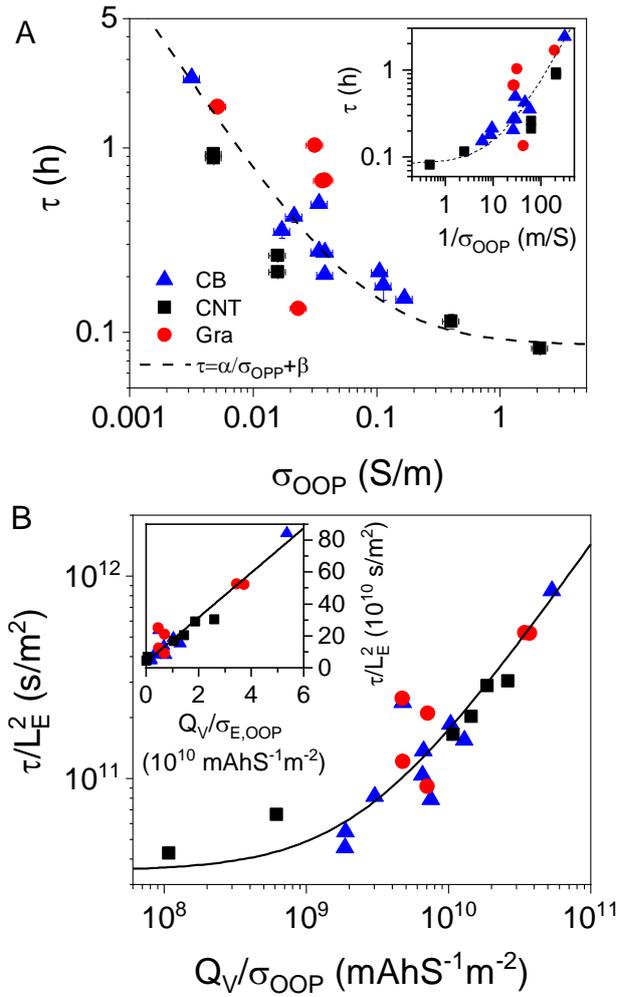

Figure 6: The effect of conductivity on rate-performance. A) Characteristic time, $\tau$, plotted versus out-of-plane electrode conductivity, $\sigma_{OOP}$, for electrodes incorporating all three conductive additives (single-walled carbon nanotubes, carbon black and graphene). The dashed line illustrates linearity between $\tau$ and $1/\sigma_{OOP}$. B) Characteristic time divided by electrode thickness squared ($\tau/L_E^2$) plotted versus low-rate volumetric capacity divided by out-of-plane electrode conductivity ($Q_V/\sigma_{OOP}$). The solid line is a plot of equation (7) using the following parameters: $\langle Q_V \rangle = 2.1\times 10^8$ mAh/m$^3$, $\sigma_{BL}$=0.5 S/m, $D_{BL}$=3×10$^{-10}$ m$^2$/s, $P_E$=0.6, $P_S$=0.4, $L_S$=16 μm, $\langle L_E \rangle$=97 μm, $L_{AM}$=r/3=2μm, $D_{AM}$=5×10$^{-14}$ m$^2$/s, $t_c$=25s (see SI for justification).

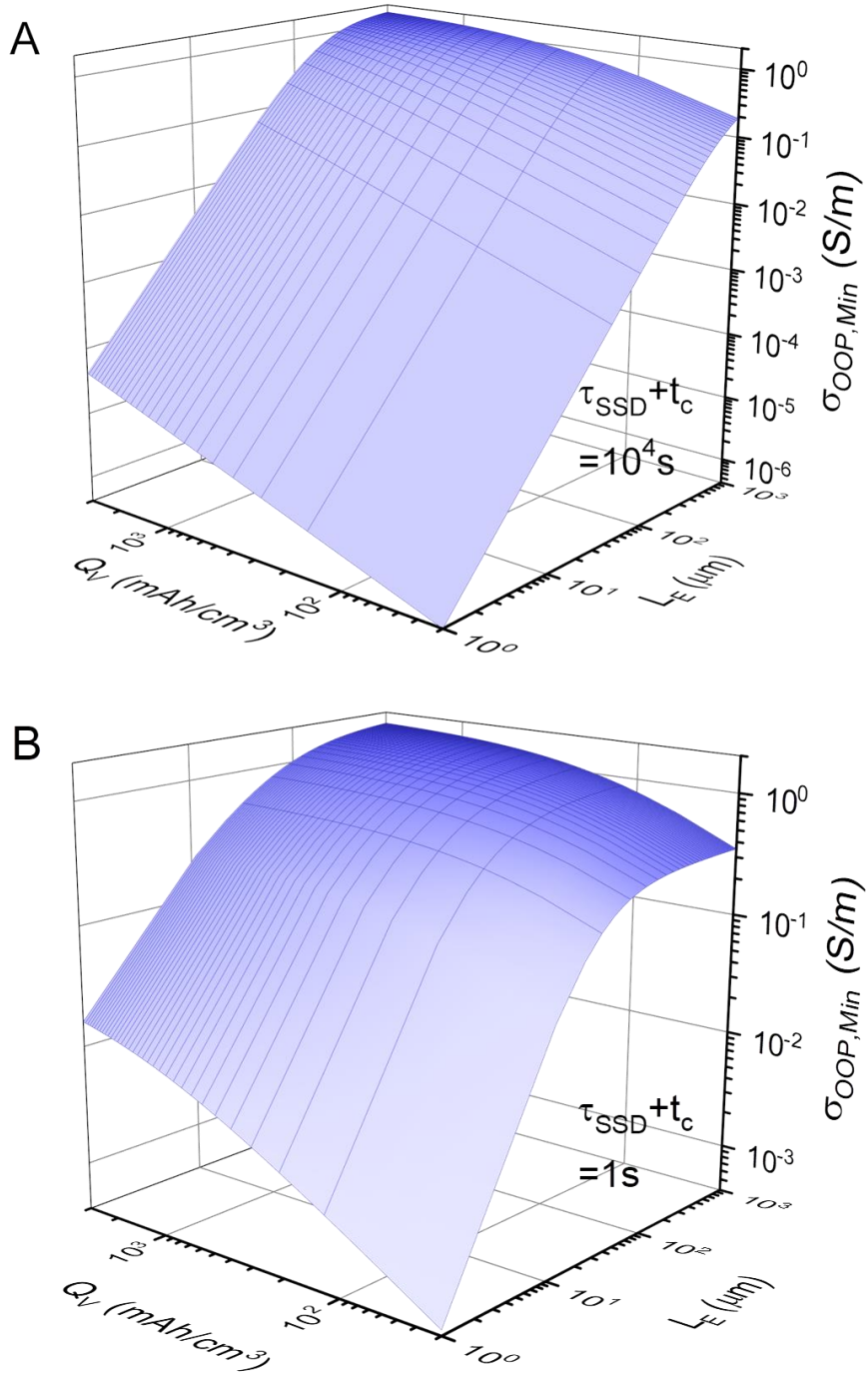

Figure 7: Critical (out-of-plane) electrode conductivity, $\sigma_{OOP,min}$, plotted as a function of electrode thickness ($L_E$) and low rate volumetric capacity ($Q_V$). The critical conductivity is that required to reduce the contribution to $\tau$ associated with the electrode resistance (first term in equation (6a)) below 10% of the sum of the other contributions to $\tau$ (i.e. the other six terms in equation (6a)). Here we calculate $\sigma_{OOP,min}$ using the following parameters: $\langle Q_V \rangle = 2.1 \times 10^8$ mAh/m$^3$, $\sigma_{BL}$=0.5 S/m, $D_{BL}$=3×10$^{-10}$ m$^2$/s, $P_E$=0.6, $P_S$=0.4, $L_S$=16 μm, $\langle L_E \rangle$=97 μm. In A and B, this calculation is performed for electrode materials with long (A, $\tau_{SSD}+t_c=10^4$s) and short (B, $\tau_{SSD}+t_c=1$s) combinations of solid-state diffusion and reaction times.

SI for

Quantifying the effect of electrical conductivity on the rate-performance of nanocomposite battery electrodes

Ruiyuan Tian,[1,2] Nolito Alcala,[3] Stephen JK O'Neill,[1,2] Dominik Horvath,[1,2] João Coelho,[2,4] Aideen Griffin,[1,2] Yan Zhang,[5] Valeria Nicolosi,[2,4] Colm O'Dwyer,[2,5] Jonathan N Coleman[1,2]*

[1]School of Physics, Trinity College Dublin, Dublin 2, Ireland

[2]AMBER Research Center, Trinity College Dublin, Dublin 2, Ireland

[3]School of Physics, Technological University Dublin, City Campus, Kevin Street, D08 NF82, Ireland

[4]School of Chemistry, Trinity College Dublin, Dublin 2, Ireland

[5]School of Chemistry, University College Cork, Tyndall National Institute, and Environmental Research Institute, Cork T12 YN60, Ireland

*colemaj@tcd.ie (Jonathan N. Coleman); Tel: +353 (0) 1 8963859.

We can plot equation 6a on figure 6B as follows.

Equation 6a can be written as:

$$\frac{\tau}{L_E^2} = \frac{14 Q_V}{\sigma_{OOP}} + \left[ \frac{28 Q_V}{2\sigma_{BL} P_E^{3/2}} + \frac{1}{D_{BL} P_E^{3/2}} + \frac{L_S}{L_E} \frac{28 Q_V}{\sigma_{BL} P_S^{3/2}} + \frac{1}{L_E^2} \left( \frac{L_S^2}{D_{BL} P_S^{3/2}} + \frac{L_{AM}^2}{D_{AM}} + t_c \right) \right]$$

When plotted as $\tau / L_E^2$ vs. $Q_V / \sigma_{OOP}$ the expected slope is 14 F/mAh while the intercept is the set of terms in the square brackets. The intercept can be found when $1/\sigma_{OOP}=0$ or when $\sigma_{OOP} \to \infty$.

For illustrative purposes, we plot $\tau / L_E^2$ vs. $\sigma_{OOP}$ below. The intercept in figure 6B is the constant value of $\tau / L_E^2$ when $\sigma_{OOP}$ becomes large. To plot this, we need to estimate the relevant parameters:

$\langle Q_V \rangle = 2.1 \times 10^8$ mAh/m$^3$     Found using $Q_V = \rho_E Q_M$ and averaging over all samples.

$\sigma_{BL} = 0.5$ S/m     Typical for LIB electrolytes[26]

$D_{BL} = 3 \times 10^{-10}$ m$^2$/s     Middle of the range for common battery electrolytes[27,28]

$P_E = 0.6$     Estimated from mean electrode density

| | |
|---|---|
| $P_S=0.4$ | Typical for commercial separators[29] |
| $L_S=16$ μm | Celgard 2032 separator |
| $\langle L_E \rangle = 97$ μm | Measured mean thickness |
| $L_{AM}=r/3 \sim 2$ μm | Proposed relationship between $L_{AM}$ and particle radius[6] |
| $D_{AM}=5\times10^{-14}$ m²/s | Diffusivity of Li ions in NMC111[30] |
| $t_c=25$ s | Roughly middle of the range reported by[6] |

Using these parameters yields the following graph which clearly shows the limiting value to be $3.5\times10^{10}$ s/m².

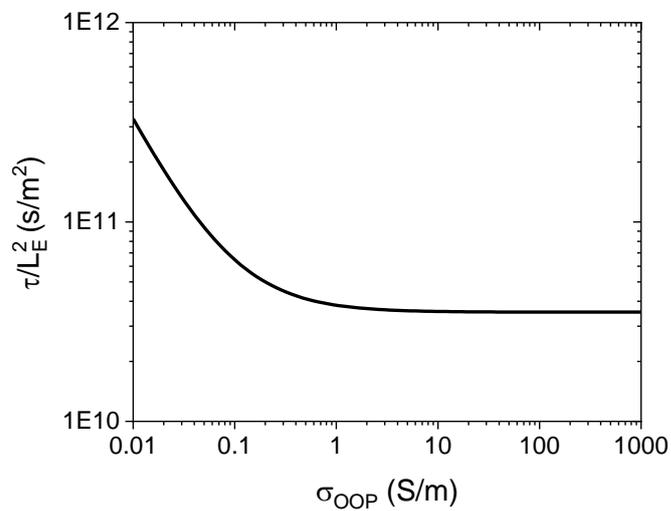